\documentclass[10pt,superscriptaddress,aps,pra,twocolumn, nofootinbib]{revtex4-2}
\usepackage{amssymb,amsmath}
\usepackage{subfig}
\usepackage{color}
\usepackage{hyperref}
\usepackage{listings}
\usepackage{physics}
\usepackage{amsmath}
\usepackage{graphicx}
\usepackage{bbold}
\usepackage{qcircuit}
\usepackage{amsfonts}
\usepackage{comment}
\usepackage{amsmath}
\usepackage{natbib}
\usepackage{multirow}
\usepackage{amsthm}

\newtheorem*{remark}{Lemma}

\usepackage[normalem]{ulem}
\usepackage[dvipsnames,usenames]{xcolor}

\hypersetup{
    unicode=false,          
    pdftoolbar=true,        
    pdfmenubar=true,        
    pdffitwindow=false,     
    pdfstartview={FitH},    
    pdfauthor={},     
    colorlinks=true,       
    linkcolor=blue,          
    citecolor=blue,        
}

\graphicspath{{figures/}}

\begin{document}

\definecolor{dkgreen}{rgb}{0,0.6,0}
\definecolor{gray}{rgb}{0.5,0.5,0.5}
\definecolor{mauve}{rgb}{0.58,0,0.82}

\lstset{frame=tb,
  	language=Matlab,
  	aboveskip=3mm,
  	belowskip=3mm,
  	showstringspaces=false,
  	columns=flexible,
  	basicstyle={\small\ttfamily},
  	numbers=none,
  	numberstyle=\tiny\color{gray},
 	keywordstyle=\color{blue},
	commentstyle=\color{dkgreen},
  	stringstyle=\color{mauve},
  	breaklines=true,
  	breakatwhitespace=true
  	tabsize=3
}

\title{Imaginary Time Propagation on a Quantum Chip}
\newcommand{\llnlafil}{Lawrence Livermore National Laboratory, P.O. Box 808, L-414, Livermore, California 94551, USA}
\newcommand{\utrentoafil}{Physics Department, University of Trento, Via Sommarive 14, I-38123 Trento, Italy}
\newcommand{\infnafil}{INFN-TIFPA Trento Institute of Fundamental Physics and Applications,  Trento, Italy}

\author{F.~Turro}
\email{francesco.turro@unitn.it}
\affiliation{\utrentoafil}
\affiliation{\infnafil}

\author{A.~Roggero}
\affiliation{\utrentoafil}
\affiliation{\infnafil}

\author{V.~Amitrano}
\affiliation{\utrentoafil}
\affiliation{\infnafil}

\author{P.~Luchi}
\affiliation{\utrentoafil}
\affiliation{\infnafil}

\author{K.~A.~Wendt}
\affiliation{\llnlafil}

\author{J.~L Dubois}
\affiliation{\llnlafil}

\author{S.~Quaglioni}
\affiliation{\llnlafil}

\author{F.~Pederiva}
\affiliation{\utrentoafil}
\affiliation{\infnafil}

\date{\today}

\begin{abstract}
Evolution in imaginary time is a prominent technique for finding the ground state of quantum many-body systems, and the heart of a number of numerical methods that have been used with great success in quantum chemistry, condensed matter and nuclear physics.    
We propose an algorithm to implement imaginary time propagation on a quantum computer.
Our algorithm is devised in the context of an efficient encoding into an optimized gate, drawing on the underlying characteristics of the quantum device, of a unitary operation in an extended Hilbert space. However, we prove that 
for simple problems it can also be successfully applied to standard digital quantum machines. 
This work paves the way for porting quantum many-body methods based on imaginary-time propagation to near-term quantum devices, enabling the future quantum simulation of the ground states of a broad class of microscopic systems.
\end{abstract}

\maketitle

\section{Introduction}
As originally proposed by Feynman in the 1980's~\cite{Feynam}, quantum computers are theorized to be  exponentially more efficient than any classical algorithm in the description of non-relativistic quantum many-body system. A notoriously hard problem for a classical computer is to find the ground state of a complex many-body system, be it in chemistry, condensed matter or nuclear physics. In most many-body methods, such as configuration-interaction or coupled-cluster,  the main limiting factor is the exponential growth of the model space with increasing number of particles or increased fidelity of the calculation.
On the other hand, in quantum Monte Carlo methods applied to Fermionic systems one has to contend with an exponential increase of the computing time with the number of particles caused by the emergence of the Fermion sign problem, which has been shown to be a NP-hard problem in general~\cite{Troyer}. 
There is therefore a desire to develop quantum versions of prominent quantum many-body methods, and in particular quantum algorithms that can be efficiently applied to emerging prototypes of quantum computing platforms, which suffer from limitations in gate error rates and quantum-device noise.

Many classical algorithms for the calculation of the ground state of microscopic systems, including projection Monte Carlo techniques, are based on the Imaginary Time Propagation (ITP) method (for an application in nuclear structure theory see e.g. Ref.~\cite{dmc_paper}). In a nutshell, this consists of solving the time-dependent Schr\"odinger equation along the imaginary time axis, rather than along the real time one (the so-called Wick rotation). The resulting evolution operator causes the exponential decay of the amplitude of all states with respect to the ground state one. Such operator can then be applied to compute the actual ground state of any given Hamiltonian starting from an arbitrary state that is not orthogonal to the ground state itself.
Methods to simulate imaginary time evolution on a quantum computer using a hybrid quantum-classical variational algorithms  were proposed in Refs.~\cite{Variationalansatz,ITP_motta}. 
Other hybrid quantum-classical approaches for computing ground states include  the variational quantum eigensolver~\cite{NatCommun54213}, which has been applied to several quantum systems~
\cite{ACSNano9,PhysRevX6,PhysRevA95,NewJPhys18,RevLett118,PRX8,PRL120,Simulationofsubatomic,Highenergy_review} and the quantum approximate optimisation algorithm proposed in Ref.~\cite{Aquantumapproximate}. In these algorithms the classical computer takes care of the optimisation of the parameters of a variational wave function while a {\em universal} quantum computer is used to perform e relatively small number of discrete qubit operations (gates) within a finite universal set. The use of such shallow quantum circuits is a common strategy to reduce the adverse impact of noise, enabling the application of the algorithm on present-day and near-term quantum hardware.

In this work we propose a method to implement ITP on a quantum computer by means of a purely quantum algorithm. 
While the real-time propagator of a quantum mechanical system is a unitary operator, the corresponding imaginary-time propagator is not, and cannot be directly translated into a quantum gate. Therefore, we devise a unitary operator that implements the ITP by working within an extended Hilbert space, given by the tensor product of the computational model space with a reservoir (or ancilla) qubit. The approach of realizing a quantum simulation by working in an extended Hilbert space  was first proposed in Ref.~\cite{Walters,Notunitary_gate_referee}, and it is at the basis of several other proposed methods for state preparation~\cite{spectralcombing,Ge2019,Near,Rog20} or Gibbs sampling~\cite{Terhal2000,Chowdhury2017,Metcalf2020}. An advantage of the technique proposed in this work over these methods is the requirement of only a single auxiliary qubit to carry out the simulation. Other algorithms, such as iterative phase estimation (see e.g.~\cite{Wiebe2016}) also share this property, but require more complex control logic to be carried out.
Different from these previous approaches, in this work we introduce an expression of the ITP that does not necessarily rely on short time propagation and  can effectively exploit a direct encoding into a single optimized gate following the optimal control (OC) approach discussed in Ref.~\cite{Optimalcontrol}.
We study the accuracy of this method in applications to the Hydrogen atom and a simple nuclear system~\cite{Optimalcontrol} by comparing an exact, classical simulation with the results obtained by implementing a circuit of primitive gates on a universal quantum computer (namely the IBM Quantum Experience system), and by simulating the implementation of optimal quantum control on the superconducting three-dimensional (3D) transmon qudit of Ref.~\cite{Supercond}. These are two of the currently available quantum computing platforms.\\
The present work is structured as follows, in Sec.~\ref{sec:quantumdmc} our quantum imaginary time propagation method is presented, in Sec.~\ref{sec:results} some results are shown. In Sec.~\ref{sec:Oneshot} there is a further discussion about ancilla probability with some improvements. In Sec.~\ref{sec:Evolution} we present an alternative to perform a sequence of imaginary time evolutions and in Sec.~\ref{sec:trotter} we provide an error analysis of this multi-step procedure and comment on the scalability of the algorithm to large systems. We finally provide our conclusions and outline future directions in Sec.~\ref{sec:conclusions}.

\section{Quantum imaginary-time propagation  \label{sec:quantumdmc} }

Before introducing the quantum version of the algorithm, here we briefly recall the basic ITP method (additional detail can be found, for example, in Ref.~\cite{dmc_paper}).

Given a time-independent Hamiltonian $H$, an arbitrary state $\ket{\psi}$ belonging to the Hilbert space of $H$ can be evolved in imaginary time $\tau = i t$ (with $t$ the real time) by formally applying the propagator $e^{-\hat{H}\tau}$ according to:
\begin{align}
\label{eq:itp}
\ket{\psi(\tau)}  = e^{-\hat H\tau} \ket{\psi(0)}
 = \sum_{n=0}^\infty c_n e^{-E_n\tau} \ket{\phi_n}\,,
\end{align} 
where we have decomposed the initial state at time $t=0$ ($\tau=0$) in terms of the eigenvectors $\ket{\phi_n}$ of the Hamiltonian, with  eigenvalues $E_n$.
We notice that the imaginary-time propagator is hermitian but not unitary. In general, this causes the normalization of the evolved state and of its components not to be preserved. It is possible to keep constant at least the normalization of the ground state, modifying the propagator in the following way:
\begin{equation}
\ket{\psi(\tau)}=e^{-(\hat{H}-E_T)\tau} \ket{\psi(0)} \label{prop}\,.
\end{equation}
The standard way is to choose $E_T$  as the ground state energy $E_0$, in this case Eq.~\eqref{eq:itp} suggests that, in imaginary time, any arbitrary state non orthogonal to the ground state $\ket{\phi_0}$ evolves to the mathematical ground state of $\hat{H}$ because the components along the excited states are exponentially suppressed. In practical applications it is not necessary to know in advance the exact value of $E_0$, but it is sufficient to have  an upperbound, which could be obtained e.g. with variational methods, with a precision $(E_T-E_0)\leq \epsilon_T$. The precise requirements for the error tolerance $\epsilon_T$ to guarantee the scheme to be stable and convergent will be provided further below.
To reproduce this algorithm on a quantum computer, we need to take into account the intrinsic non-unitarity of the process. One possibility to implement a dissipative process is to extend the Hilbert space coupling it to a reservoir, and transfer probability to the components of the computational basis orthogonal to the original Hilbert space. Let us map an arbitrary state of the physical system onto a (multi)qubit state $\ket{\psi_s}$. We then introduce the product state of $\ket{\psi_s}$ with a reservoir qubit prepared in the $\ket{0}$ state, yielding the total initial wave function:

\begin{equation}
\ket{\Psi_{\rm init}}= \ket{0} \otimes \ket{\psi_s} =
\left(\begin{array}{c} 1 \\ 0\\
\end{array}\right) \otimes  \ket{\psi_s} \,,  
\label{initial}
\end{equation}

and the unitary operator acting on the total Hilbert space (system times reservoir qubit)

\begin{equation}
\hat{U}(\tau)=\begin{pmatrix}
\hat{Q}_{\rm ITP}(\tau) & \frac{1}{\sqrt{\mathbb{1}+e^{-2\left(\hat H-E_T\right)\tau}}}\\
\frac{1}{\sqrt{\mathbb{1}+e^{-2\left(\hat H-E_T\right)\tau}}}&-\hat{Q}_{\rm ITP}(\tau)\\
\end{pmatrix}\;. \label{eq:Utau}
\end{equation}

Each matrix element in Eq.~\eqref{eq:Utau} is an operator acting on the Hilbert space describing the physical system, with $\hat Q_{\rm ITP}(\tau)=\left(\mathbb{1}+e^{-2\left(\hat H-E_T\right)\tau}\right)^{-1/2} e^{-\left(\hat H-E_T\right)\tau}$ and $\mathbb{1}$ being, respectively, the normalized imaginary-time propagator and the identity operator . (A proof that $\hat{U}(\tau)$ is a unitary operator is provided in Appendix~\ref{supp_unitary}). 
Notice that $\hat{U}(\tau)$ can also be written as the sum of the tensor products $\hat\sigma_z\otimes \hat Q_{\rm ITP}(\tau)$ and  $\hat\sigma_x \otimes \left(\mathbb{1}+e^{-2\left(\hat H-E_T\right)\tau}\right)^{-1/2} $, where $\hat\sigma_z$ and $\hat\sigma_x$ are the Pauli $Z$ and $X$ operators acting on the reservoir qubit.

The application of $\hat U(\tau)$ to $\ket{\Psi_{\rm init}}$ yields:

\begin{align}
\label{eq:initial}
    \ket{\Psi(\tau)} ~=&~~\ket{0}\otimes \hat Q_{\rm ITP}(\tau) \ket{\psi_s} \\\nonumber
    & \!\!+ \ket{1} \otimes  \left(\mathbb{1}+e^{-2\left(\hat H-E_T\right)\tau}\right)^{\!-\tfrac{1}{2}}\ket{\psi_s}\,.
\end{align}

It follows that, performing a measurement along the reservoir state $\ket{0}$, the total system collapses to the state:

\begin{align}
    \ket{\Psi_{\rm fin}}=C \ket{0} \otimes \hat Q_{\rm ITP}(\tau)\ket{\psi_s}\,,
    \label{res_fin}
\end{align}

where $C$ is a coefficient introduced to normalize the state after the measurement. The expression in Eq.~\eqref{res_fin} is analogous to the standard imaginary time propagator applied to the initial wave function of Eq.~\eqref{prop}. There are however some important differences in the quantum ITP scheme. 
Above all, in the quantum algorithm version of ITP, the normalization constant $C$ is of great importance as it's square gives the success probability of the imaginary-time step (i.e. the probability of measuring the reservoir qubit in the $\ket{0}$ state after the application of $\hat{U}(\tau)$). Fortunately we can ensure that the success probability $P(0)=|C|^2$ is reasonably larger than $0$ for any value of the imaginary-time step $\tau$. It is in fact easy to show (see Appendix~\ref{app:sprob_proof} for a full derivation) that, if we indicate with $c_0=\langle\phi_0|\psi_s\rangle$ the initial overlap with the ground-state, one has
\begin{equation}
\label{eq:psucc_low_bound}
P(0) \geq \frac{|c_0|^2}{1+\exp\left(2\tau(E_0-E_T)\right)}\;.
\end{equation}

\begin{figure*}[th]
$$\Qcircuit @C=1.5em @R=1em {
   \ustick{q_0}& \qw & \gate{U_3(2.8593,-\frac{\pi}{2},0)}  & \ctrl{1} &  \gate{U_3(\pi,-0.0278,4.6846)} & \ctrl{1} &  \gate{U_3(0.2823,0,-\pi)} & \qw\\
 \ustick{q_1}&  \qw & \gate{U_3(\frac{\pi}{2},0.2823,\frac{3\pi}{2})}  & \targ &  \gate{U_3(\frac{\pi}{2},\frac{3\pi}{2},0.7854)} & \targ &  \gate{U_3(0.5031,0,0)} & \qw\\
 } $$
 
 \caption{Gate set for implementing $U(\tau)$ for the Hydrogen atom Hamiltonian with $\tau=15$ Hartee$^{-1}$. $q_1$ represents the reservoir qubit and $q_0$ the system qubit.\label{gate_Hyl}}
\end{figure*}
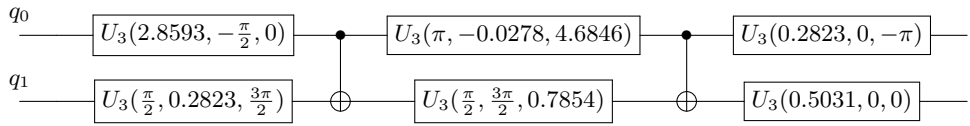

This shows that if, as indicated above, we take $E_T\geq E_0$ to be an upperbound on the ground-state energy then $P(0)\geq |c_0|^2/2$ even at long imaginary times (note that, if $E_T$ is strictly greater than $E_0$, this bound actually gets better converging to $P(0)\geq |c_0|^2$ as $\tau\to\infty$). On the other hand, if we take $E_T<E_0$ instead, $P(0)$ decays exponentially to $0$ with imaginary-time, preventing an efficient use of the quantum ITP algorithm. Obtaining a large value for $P(0)$ is critical for the multi-step (short-time propagation) procedure that will be described in Sec.~\ref{sec:Evolution} since the final success probability after $M$ steps decays exponentially as $\approx P(0)^M$. One method to achieve large $P(0)$ is clearly to improve the initial state fidelity (i.e. increase $|c_0|^2$) by optimizing the starting state $\ket{\psi_s}$. In addition, it is possible to effectively increase $P(0)$ to values close to one by using amplitude amplification and increasing the depth of the quantum circuit by a factor $\approx1/\sqrt{P(0)}$~\cite{amplitude_amplification}. We will show results obtained on real quantum hardware for the latter approach in Sec.~\ref{sec:Oneshot}. 

More generally, one can prove that $\ket{\Psi_s(\tau)}$is closer to the ground state of $\hat{H}$ than the initial state (see proof in Appendix~\ref{app:gfid_proof}). 
In fact, the overall effect of $\hat Q_{\rm ITP}(\tau)$ can be computed in two important limits: 

\begin{itemize}

\item The limit for  $\tau\, \!\!\!\rightarrow\!\!\!\, 0$. In this limit, 
expanding at first order in $\tau$, one can show that:

$\left(\mathbb{1}+e^{-2\left(\hat H-E_T\right)\tau}\right)^{-1/2}\simeq \frac{1}{\sqrt{2}}e^{\left(\hat H-E_T\right)\tau/2}$, 

and hence the action of $\hat Q_{\rm ITP}(\tau)$ corresponds to the application of a classical ITP algorithm, but over an imaginary-time interval $\frac{\tau}{2}$.  This limit is useful when adopting a Trotter-Suzuki decomposition of the Hamiltonian.

\item The limit for  $\tau\!\gg\! \frac{1}{E_1-E_0 }$, where $E_1$ is the energy of the first excited state and $E_0$ is the ground state energy. In this limit, 
the ground state $\ket{\phi_0}$ of the system can be obtained with a single application of the operator $\hat Q_{\rm ITP}$ choosing $E_T < E_1$ (see Appendix~\ref{app:gfid_proof}). 
In the case $E_T=E_0$, it follows that $\hat{Q}_{\rm ITP}(\tau)\ket{\psi_s}\xrightarrow[]{\tau \rightarrow \infty} \frac{c_0}{\sqrt{2 }} \ket{\phi_0}$  where $c_0=\braket{\phi_0}{\psi_s}$.
\end{itemize}

One can also observe that, after the measurements, the final state of the system and reservoir qubits are in the same initial condition as in Eq.~\eqref{initial}. Therefore, repeating this algorithm $N$ times yields the evolution for a total imaginary time $\tau_{\rm tot}>N\,\frac{\tau}{2}$ for a generic time step $\tau$.\\

\section{Results \label{sec:results}}
As a demonstration, we applied our quantum ITP algorithm to two physical problems: finding the ground state of the Hydrogen atom,  and the search of the lowest energy spin state of two neutrons at a given distance, as described in Ref.~\cite{Optimalcontrol}. For the Hydrogen atom, we implemented 
two different strategies, i.e.\ expressing the unitary transformation in terms of standard quantum gates (namely, $R_x$, $R_z$ and $CNOT$, as shown in Fig.~\ref{gate_Hyl}, in which  $U_3$ can be decomposed in multiplication of $R_x$ and $R_y$)  and the optimal control approach for a trasmon qudit as discussed in Ref.~\cite{Optimalcontrol}. For the two neutron system, we opted for the optimal control method only. Additional details can be found in Appendix~\ref{supp:opt}. Experimentally, the occupation probability of each state in the computational basis when the system has reached the ground state,  $p_\beta(\tau_{\rm tot})=|\langle\beta|\psi_s(\tau_{\rm tot})\rangle|^2$ with $\ket{\beta}=\ket{0},\ket{1},...$, can be obtained  from the occupancies measured when the reservoir qubit is measured in the $\ket{0}$ state, $p_{0\beta}$, according to:

\begin{equation}
\label{normal}    
p_\beta(\tau_{\rm tot})=\frac{p_{0\beta}(\tau_{\rm tot})}{\sum_{i\in \{\beta\}} p_{0i}(\tau_{\rm tot})}\,,
\end{equation}
where $\beta$ indicates the index of the physical state (for n qubit $\beta$ runs from 0 to $2^n-1$). In our words, we normalize the state \\
\label{sec:Hy}

In the simulation of the Hydrogen atom, we assume spherical symmetry in the s-wave and expand the radial wave function on the STO-2G basis, i.e.\ a linear combination of two Gaussian functions that approximate a Slater-type orbital~\cite{STO-NG}. 
Because this is a non-orthogonal basis, before applying our quantum algorithm, we first orthonormalize the Hamiltonian matrix (see, e.g., Eq.~(5.15) of Ref.~\cite{Not_orthogonal_basis_set}.
The quantum ITP algorithm requires two qubits: one qubit to represent the two orthonormalized basis states plus an additional reservoir qubit.
We prepare the system-plus-reservoir wave function in the initial condition of Eq.~\eqref{initial} in the $\ket{x}$ state ($|c_0|^2=0.361$), and apply our quantum ITP algorithm in a `single shot' (i.e., working in the $\tau\gg 1$ limit) using $\tau_{\rm tot}=\tau=15.0$ Hartee$^{-1}$.

\begin{figure}[t]
\includegraphics[scale=0.2]{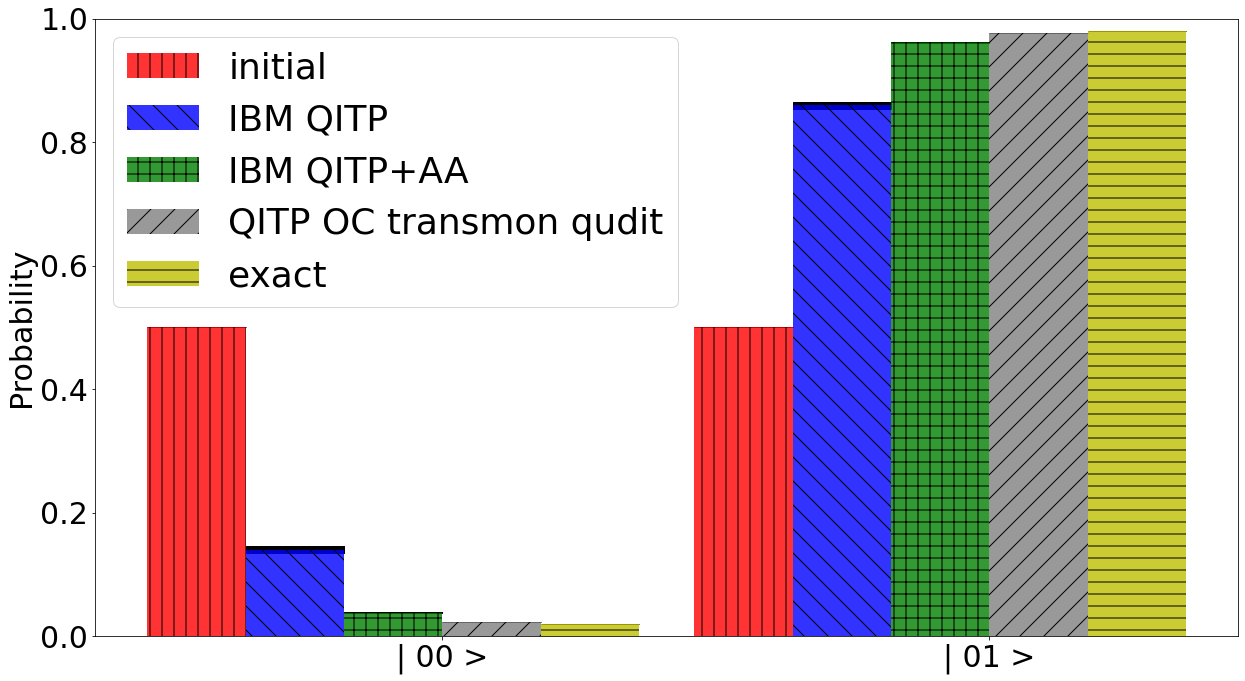}
\caption{Normalized occupation probabilities ($p_\beta$) for the computed wave function of the Hydrogen atom using the STO-2G basis set at $\tau=0$ (red bar) and $\tau=15.0$ Hartee$^{-1}$ in two different approaches: the IBM result (blue and green bars) and the optimal control with   inclusion of hardware noise (grey bar) compared to the exact ground-state distribution (orange bar). The black bar represents the uncertainties. The blue IBM  results are obtained applied the QITP algorithm and the green results applied the QITP operator with the amplitude amplification (AA) algorithm \label{fig:dmc_Hy}} 
\end{figure}
Fig.~\ref{fig:dmc_Hy} shows the simulated normalized occupation probabilities $p_{\beta}(\tau_{\rm tot})$ of the two computational states. 
The results obtained by running the quantum circuit of Fig.~\ref{gate_Hyl} on the free access IBM \textit{ibmq$\_$santiago} system~\cite{IBM_quantum_exp} with 8192 shots are compared with a device level simulation of the implementation on a generic multilevel transmon using the optimal control approach as in Ref.~\cite{Optimalcontrol}. We note that all the results reported here do not employ any form of error mitigation but are instead the bare outcomes of either an experiment (for the IBM device) or a classical simulation (for the OC results). The possible increase in logical fidelity afforded by error mitigation will be explored in detail in future work. 
We report in Table~\ref{tab:fidelity} the obtained fidelities between the normalized final state and  the true ground state. For the state identification, we used the method illustrated in Appendix~\ref{app:tomography}.\\
One can see that the final state is essentially the ground state for the optimal control simulation and quite close to it in the IBM results. We want to highlight that the results of IBM are obtained on a real device, instead, the OC data are the result fo a classical device-level simulation of the transmon qudit obtained by solving the Master equation with a realistic noise model.
Using the OC approach one expects to reduce the contribution of gate infidelity in the results and this will be put to test with an implementation on the physical device in a later study.

\begin{table}[ht]
\begin{tabular}{|c|c|}

\hline
\multicolumn{2}{|c|}{Hydrogen atom system}\\
\hline
Simulation & Fidelity\\
\hline
Simulator of 
transmon qudit with noise$^2$ & 0.9978\\
Fidelity 
transmon qudit without noise$^2$ &  1.00000000\\
IBM ITP$^1$ &   0.942(4)\\
IBM ITP+AA$^1$ & 0.9968(4)\\

\hline
\multicolumn{2}{|c|}{Nuclear spin system}\\
\hline
Simulation & Fidelity\\
\hline
Simulator of 
transmon qudit with noise$^2$ & 0.9919\\
Fidelity 
trasmon qudit without noise$^2$ & 0.999999996\\
\hline
\end{tabular}
    \caption{Fidelity results for the different applications of ITP algorithm. The $^1$ indicates the fidelity is computed with the density matrix obtained solving the Master equation; the $^2$ indicates the fidelity is computed with the state estimated with the tomography described in Apppendix~\ref{app:tomography} }
    \label{tab:fidelity}
\end{table}

As a second example, we apply our quantum ITP algorithm to a system of two neutrons interacting with a chiral effective field theory ($\chi$-EFT) nucleon-nucleon potential at leading order (LO) ~\cite{37,38}, characterized by a spin-state dependence that includes full tensorial terms.
A detailed expression for the Hamiltonian is given in Appendix~\ref{supp:nuclear}. 
As previously done in Ref.~\cite{Optimalcontrol}, we consider the simplified case in which the two neutrons are `frozen' in space, reducing the problem to the description of two spins interacting through the nuclear Hamiltonian at a fixed separation.

\begin{figure}[t]
\includegraphics[scale=0.2]{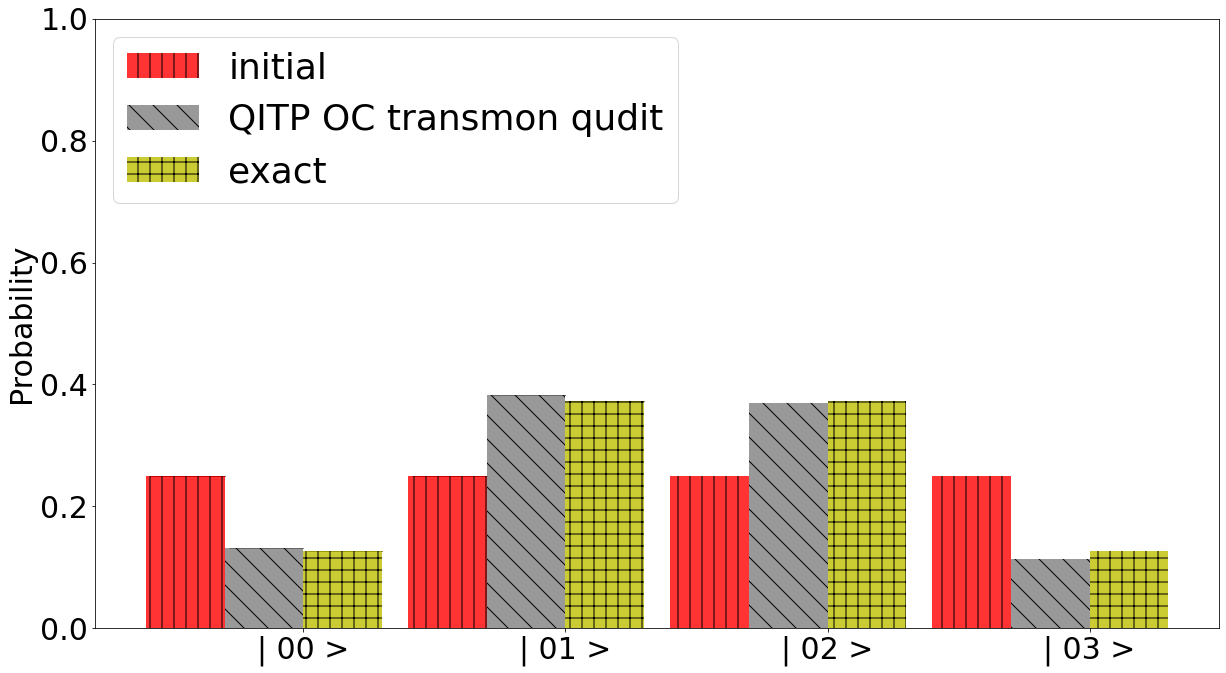}
\caption{Normalized occupation probabilities $(p_{\beta})$ for the two spin neutron system with $\tau=1$ MeV$^{-1}$ in the optimal control approach. Same legend of Fig.~\ref{fig:dmc_Hy}. \label{fig_dmcnn}}
\end{figure}

In this example, the extended Hilbert space is covered by the first 8 levels (3 qubits) of a qudit, representing the tensor product between the 4 levels (2 qubits) used to describe the spin state of the two neutrons, and the two levels of the reservoir qubit. The mapping of the computational states to the quantum processor is as follows: the $\ket{0}$ Fock state corresponds to the tensor product of the reservoir qubit in state $\ket{0}$ with the uncoupled spin state of $\ket{0}\otimes\ket{\downarrow\downarrow}$. Likewise, $\ket{1}$, $\ket{2}$, and $\ket{3}$ correspond, respectively, to the tensor products with the uncoupled spin states $\ket{0}\otimes\ket{\downarrow\uparrow}$, $\ket{0}\otimes\ket{\uparrow\downarrow}$, and $\ket{0}\otimes\ket{\uparrow\uparrow}$. The next four Fock states have a similar mapping, except for the reservoir qubit being in the $\ket{1}$ state.
The system is initially prepared in the state $\ket{\psi_s(\tau=0)}=\frac{1}{2}{\left( \ket{0}+\ket{1}+\ket{2}+\ket{3}\right)}$ with overlap $|c_0|^2=0.407$. As before, we work in the one-shot limit, $\tau \gg 1/(E_1-E_0)$, in particular, we simulated a single propagation over an imaginary time  $\tau_{\rm tot}=\tau=1$ MeV$^{-1}$. In Fig.~\ref{fig_dmcnn} we present the  simulated occupation probabilities of the two neutrons spin states. Once again, the probability distribution is already very close to that of the exact ground state. The fidelity is shown in Table~\ref{tab:fidelity}.\\
The robustness of the algorithm is discussed at the end of Appendix~\ref{app:gfid_proof}, where we also show examples on how the final state depends on the choice of the parameter $E_T$ in the propagator.

\section{Ancilla probability vs parameters of the systems \label{sec:Oneshot}}
When one applies the imaginary time method, the evolution depends mainly on the energy shift $E_T$, the time step $\tau$ and $c_0$, the overlap between the initial state and the ground state (GS). 
Following the asymptotic limit presented in Sec.~\ref{sec:quantumdmc} one can easily  verify the ancilla probability in the $\ket{0}$ state tends to $\frac{1}{2}$ in the limit $\tau \xrightarrow[]{} 0$. 
Instead, in the limit $\tau \xrightarrow[]{} \infty$ it goes to $\frac{\left|c_0 \right|^2}{1+e^{2\,\tau\,(E_0-E_T)}}$.

One can prove that for a generic time step $\tau$ the probability of measuring the ancilla qubit in the $\ket{0}$ state  is between $\frac{|c_0|^2}{2}$ and $\frac{\left|c_0 \right|^2}{1+e^{2\,\tau\,(E_0-E_T)}}$. For instance, choosing $E_T=E_0$, it is in the interval $\left[\frac{|c_0|^2}{2},\frac{1}{2}\right]$. Choosing $E_T>E_0$ one can increase the success probability in the long time limit up to $|c_0|^2$. 
More details can be found in Appendix~\ref{app:sprob_proof}.
To increase the probability of success of this quantum algorithm, that is to the probability to measure the $\ket{0}$ state of the ancilla,  one can use quantum amplitude amplification~\cite{amplitude_amplification}.

\begin{figure*}[bht]
    \centering
    \includegraphics[scale=0.25]{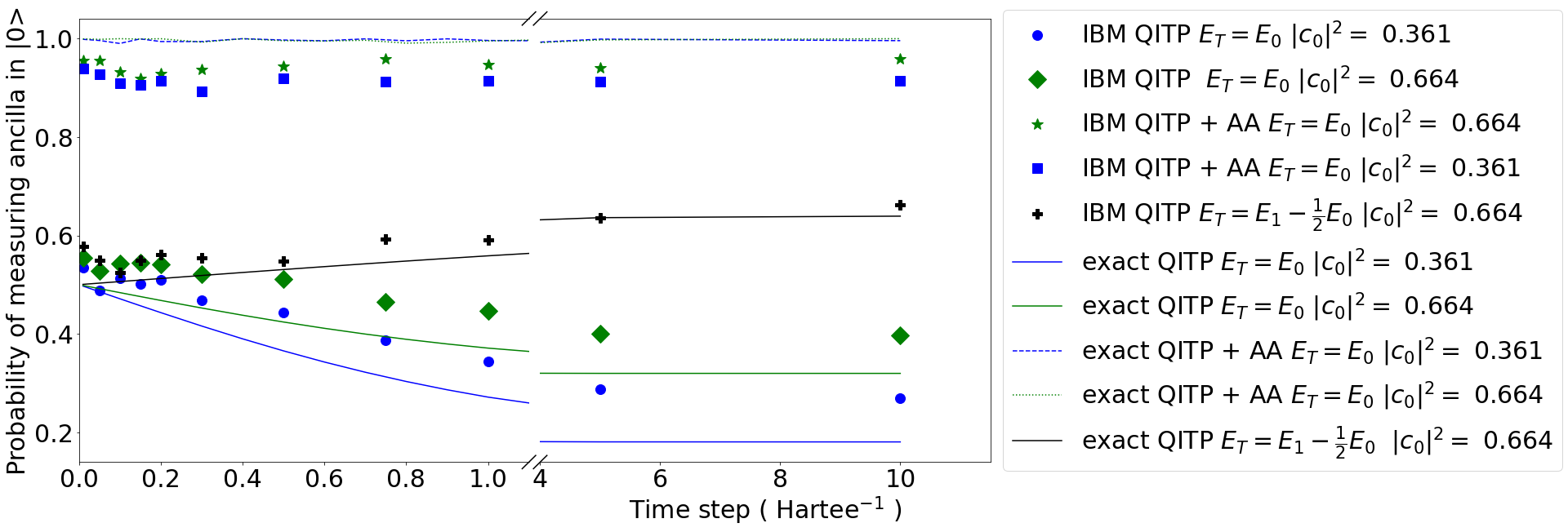}
    \caption{Ancilla probability in the $\ket{0}$ state as a function of time step $\tau$ and for different overlap with the ground state (GS) applying the ITP operator and ITP operator with the amplitude amplification(AA) with different values of the trial energy $E_T$. All uncertainties are smaller than $5\;10^{-3}$.
    }
    \label{fig:ancilla_probability}
\end{figure*}

In Fig.~\ref{fig:ancilla_probability} we present the obtained results  for ancilla probability in the $\ket{0}$ state as a function of the imaginary time step for different initial state. The squares and stars indicate the data from the \textit{ibmq$\_$manila} QPU  for a single application of the QITP  algorithm, the circles and diamonds represent the data running on IBM adding the amplitude amplification after the bare QITP. The blue circles and squares are computed from an initial state with an overlap probability of $0.361$ with the GS and the green diamonds and stars one with $0.639$. The dashed and continuous lines represent the exact results. Applying amplitude amplification there is a great improvement of measuring the ancilla probability in $\ket{0}$. Moreover, the experimental and exact results for raw ITP using a trial energy greater than the ground energy, in this case  $E_T=E_0+\frac{1}{2}E_1$, are represented by the black plus symbols and line respectively. One can observe a raising of the success probability of the ITP algorithm.

\section{Imaginary Time Evolution\label{sec:Evolution}}

The application of a sequence of short-time imaginary time propagators may lead to a more practical (and possibly scalable) algorithm  because in the limit $\tau\rightarrow 0$ it may be possible to generate accurate approximations to the $\hat{U}(\tau)$ unitary of Eq.~\eqref{eq:Utau} for large qubit systems. A drawback of the present method is the requirement of measuring the ancilla qubit in $\ket{0}$ state at each time step to continue towards the ground state. This can slow down the efficiency of the method. Luckily, using the amplitude amplification method at each time step one can avoid measuring the ancilla qubit because it would be automatically in the initial condition of the present QITP method. In addition, one can also avoid measuring the reservoir qubit at each iteration by applying the same gates with different reservoir qubits, and easily reach a long time propagation.
In the event the quantum circuit should become very deep, one can measure the reservoir qubits after a number of steps and perform a complete state tomography to identify which state has been obtained and then re-inizialize the quantum circuit with the evaluated state. This is similar in spirit to the ``restarting" procedure proposed in ~\cite{otten2019noiseresilient},  where the state after one step of dynamical evolution is approximated by optimizing a variational circuit before performing the next time step. In general, the state tomography procedure proposed here is not in  scalable to more than a couple of qubits for the system register, but can provide interesting benchmarks on small scale near-term devices.

\begin{figure*}[ht]
    \centering
    \includegraphics[scale=0.25]{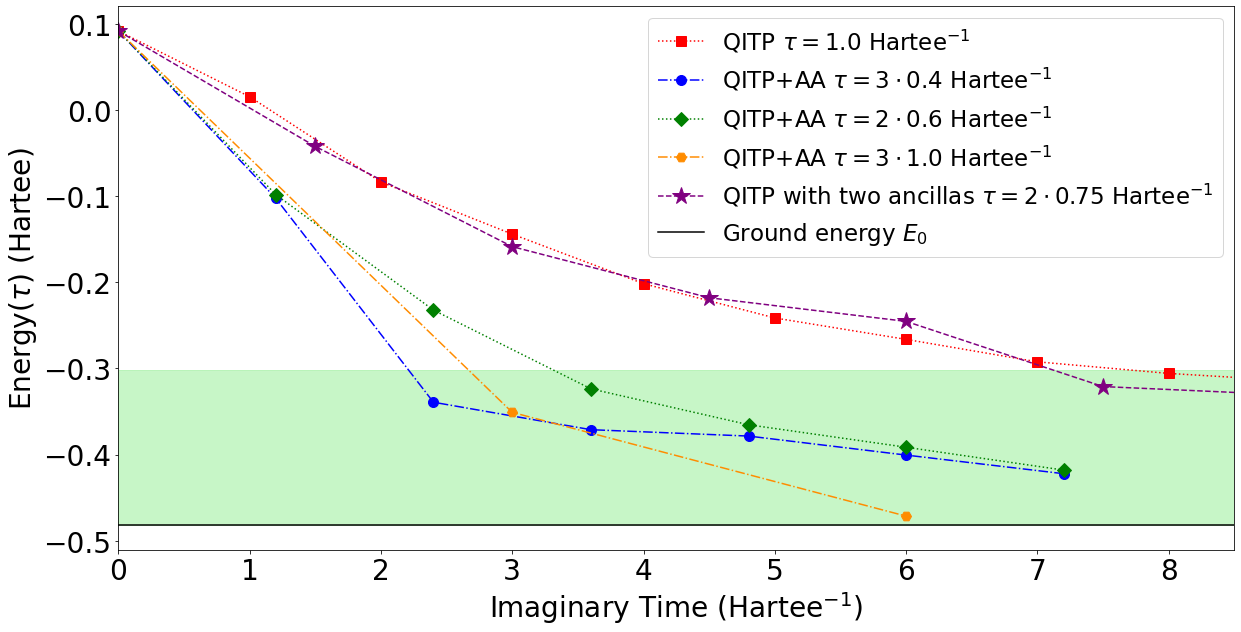}
    \caption{Results for  the expectation value of the energy as a function of imaginary time for different time steps $\tau$ with a re-initialization procedure. The light-green color bar indicates the energy range with fidelity with GS greater than $80\%$ .The cases indicated with AA are computed with Amplitude Amplification. The case labeled with ``2nd ancilla" represented the simulation done with 2 ancilla qubits. The integer factor in front of the time step indicates the number of repetitions of the ITP quantum circuit before the tomography procedure. 
    } 
    \label{fig:resultsteps}
\end{figure*}

An example of imaginary-time evolution is shown in Fig.~\ref{fig:resultsteps} for the Hydrogen atom using the $manila$ IBM QPU. 
At each time step, we initialize the state with a  $U_3$ gate on the state evaluated in the previous round, allowing for a repeated application and for a propagation over a large total imaginary time. To establish in which state the circuit would be found at the end of each step, one  can perform three measurements, i.e. the probability for the bare circuit, that for the circuit plus an additional $R_x(-\frac{\pi}{4})$ gate, and that for an additional $R_y(-\frac{\pi}{4})$ gate. From these rotations one can determine the relative phase between the two states of the system qubit (see Appendix~\ref{app:tomography}
for details). Since we are constraining the results to be described by a pure state, this tomography procedure performs a purification of the system's qubit.
Figure~\ref{fig:resultsteps} shows the energy at each iteration.
The result for $\tau=1.0$ is computed with a single application of $\hat{U}(\tau=1.0 \;\text{Hartee}^{-1})$. For $\tau=2\cdot0.75$, instead, we apply twice the $\hat{U}(\tau=0.75\,\text{Hartee}^{-1})$ with two different ancilla qubits. The other results are computed using the Amplitude Amplification (AA) algorithms where the integer factor multiplying the time step indicates the number of repetitions of ITP(+AA) quantum circuit before the tomography procedure. 
The shaded area shows the energy of states that reach the GS with fidelity greater than $80\%$. The time shown in the plot is the time parameter used in the $Q_{ITP}(\tau)$ propagator which, in general, cannot be directly matched to the time $\tau$ in the standard ITP propagator  $e^{-\tau (H-E_T)}$ due to the presence of the normalization factor. At the end of the imaginary time evolution, all quantum simulations converge to a state very close to the GS.\\

\section{Trotter decomposition}
\label{sec:trotter}

In order to understand the prospects for scalability of the QITP algorithm for large system size, in this section we present results for approximating the QITP unitary using a Trotter-like decomposition of the unitary $\hat{U}(\tau)$ from Eq.~\eqref{eq:Utau}. For Hamiltonians that can be written as $\hat{H}=\sum_{l}^L \hat{H}_l$, with $\hat{H}_l$ acting only on a constant number of qubits, the scheme consists in applying $r$ times a short-time unitary $\hat{U}(\delta \tau)$ with $\delta\tau=\tau/r$ and then approximate each operation as a product of $L$ unitaries. The post-selection in the $\ket{0}$ state of the ancilla used to define the individual unitary (cf. Eq.~\eqref{eq:Utau}) can be performed by either using $Lr$ ancilla and post-selecting on their state being in $\ket{0}^{\otimes Lr}$ or, equivalently, using a single ancilla qubit and post-selecting it for $Lr$ times. We start by describing the decomposition of the full evolution into $r$ short-time steps, and how the fidelity is affected by the decomposition. We then show how to approximate each short-time step with controllable error and close this section by showing bound on the total success probability. The full operator obtained after time-stepping can be written as
\begin{equation}
\label{eq:rep_op}
    \left(\hat{M}_0 \hat{U}(\delta\tau) \right)^r = \begin{pmatrix}
    \left( \frac{e^{-\delta\tau (\hat{H}-E_T)}}{\sqrt{\mathbb{1}+e^{-2\delta\tau(\hat{H}-E_T)} }}\right)^r & 0 \\
    0 & 0\\
    \end{pmatrix}\,,
\end{equation}
where $\hat{M}_0=\ket{0}\bra{0}$ is the projector operator to the $\ket{0}$ in the ancilla qubit. Using this sequence of operations the fidelity with the ground-state becomes then
\begin{equation}
    F= \left(1+\sum_n \frac{|c_n|^2}{|c_0|^2} \frac{e^{-2\tau(E_n-E_T)}\left(1+e^{-2\,\delta\tau\,(E_0-E_T)}\right)^r}{e^{-2\tau(E_0-E_T)} \left(e^{-2\,\delta\tau(E_n-E_T)}+1 \right)^r } \right)^{-1}
\end{equation}
Using the result from Eq.~\ref{eq:bound_fidelityGS} in Appendix~\ref{app:gfid_proof} for the fidelity bound and using $r\in\mathbb{N}$ we get:
\begin{equation}
\label{eq:fidelity_bound_r}
F\geq\left(1+\frac{1-|c_0|^2}{|c_0|^2}\left(\frac{e^{2\delta\tau(E_0-E_T)}+1}{e^{2\delta\tau(E_1-E_T)}+1}\right)^r\right)^{-1}\;,
\end{equation}
Assuming $E_1>E_T>E_0$ and considering small time-steps $2\delta\tau\,\Delta\,<\,1$, one can prove that 
\begin{equation}
\label{eq:bound_ratio}
\left(\frac{e^{2\delta\tau(E_0-E_T)}+1}{e^{2\delta\tau(E_1-E_T)}+1}\right)^r
\,\leq\,\exp\left(-\frac{1}{2}r\delta\tau\Delta\right)\;.
\end{equation}

For a derivation of this result see Eq.~\eqref{eq:fid_fac_bound} in Appendix~\ref{app:Trotter_proof}. This shows that the fidelity converges exponentially to one in $r\delta\tau\Delta$. In general this will however still be subexponential in the number of steps $r$ since we need to decrease $\delta\tau$ with $r$ in order to control the approximation error. To see this we use the following Lemma.
\begin{remark}
Let a Hermitian operator $\hat{H}$ be expressed as a sum of $L$ Hermitian operators $\{\hat{H}_1,..,\hat{H}_L\}$, $\hat{H}=\sum_l^L \hat{H}_l$ and $\Lambda=\,\max_j\,\left\|\hat{H}_j-E_T\right\|_{\infty}$. Then, 
\begin{equation}
\label{eq:trotter_bound_main}
\left\|\hat{Q}_{ITP}(\tau)-\prod_k^L\hat{Q}^{(k)}_{ITP}(\tau)\right\|\leq L^2\Lambda_T^2 \tau^2\;,
\end{equation} 
where $\hat{Q}_{ITP}^{(k)}(\tau)\,=\,  \frac{e^{-\tau \left(\hat{H}_k-\frac{E_T}{L}\right)}}{\sqrt{1+e^{-2\tau\left(\hat{H}_k-\frac{E_T}{L}\right)} }} $.
\end{remark}
This result can be obtained using similar techniques to those in Appendix F of Ref.~\cite{Trotter_error}. We provide a detailed proof and a generalization of this result in Appendix~\ref{app:Trotter_proof}. The key advantage of this strategy is that, if the Hamiltonians in the decomposition are at most $m$-local (ie. they act non-trivially on at most $m$ quibts) their synthesis using standard universal gate sets, composed of one- and two-qubit operations, can be achieved with a gate count scaling as $\mathcal{O}(4^k)$ (see e.g.~\cite{multiqubit2004}). For small $k$ one can also use Optimal Control strategies like those described in Appendix~\ref{supp:opt}.
In order to use this to approximate the $r$ steps with total error $\epsilon$ we then need
\begin{equation}
\delta\tau \leq \sqrt{\frac{\epsilon}{r}}\frac{1}{L\Lambda_T}\ll1\;.
\end{equation}
This result can be obtained applying the union bound on the product of $r$ Trotterized evolution and using Eq.~\eqref{eq:trotter_bound_main}. Using this bound, together with the results in Eq.~\eqref{eq:fidelity_bound_r} and Eq.~\eqref{eq:bound_ratio}, this shows that the fidelity with the ground-state goes to $1$ as:
\begin{equation}
 F\,\le \,\frac{1}{1+\frac{1-|c_0|^2}{|c_0|^2}\,\mathcal{O}\left(\exp\left(-\frac{1}{2}\frac{\Delta}{L\Lambda_T}\sqrt{r\epsilon}\right)\right) }\,.   
\end{equation}

The proof and more details are shown in App.~\ref{app:Trotter_proof}. This result shows that the Trotterized QITP algorithm converges super-polynomially but sub-exponentially to unit fidelity with the ground state as we increase the number of steps $r$. The scaling $\exp(-\mathcal{O}(\sqrt{r}))$ can possibly be improved using higher order Trotter-Suzuki formulas~\cite{Suzuki91} (see also~\cite{Trotter_error}). 

Finally we need to assess the behavior of the success probability for the scheme proposed above. We start discussing the bound on the probability to obtain the operator in Eq.~\eqref{eq:rep_op} using the exact short-time QITP step. Using the result found in Appendix~\ref{app:sprob_proof} we have
\begin{equation}
P(0;r)\geq \frac{|c_0|^2}{(e^{2\delta\tau(E_0-E_T)}+1)^r}
\end{equation}
which, for the choice $E_1>E_T>E_0$ can be rewritten more clearly as
\begin{equation}
\begin{split}
\label{eq:sprob_full}
P(0;r)\geq \frac{|c_0|^2}{(e^{-2\delta\tau|E_0-E_T|}+1)^r}\geq \frac{|c_0|^2}{2^r}
\end{split}
\end{equation}
which decays exponentially with the number of steps.
In Appendix~\ref{app:Trotter_proof} we present a generalization of the unitary $\hat{U}(\tau)$ with an additional free parameter that can be tuned to improve this scaling at the expense of a slower convergence of the fidelity with step number $r$. For the full scheme using the approximation in Eq.~\eqref{eq:trotter_bound_main} for $r$ steps with a total error bounded by $\epsilon$, the success probability is reduced by at most an additive factor $2\epsilon$ (see Appendix~\ref{app:Trotter_proof} for a derivation). Since the bound in Eq.~\eqref{eq:sprob_full} becomes exponentially small, this indicates that there is a maximum number of steps for which we can guarantee a non-zero success probability. This is possibly a problem with the lower bound more than with the algorithm itself and in future work it will be useful to both benchmark this result on model systems but also to derive upper-bounds on the success probability.
Finally, using the first order Trotter expansion discussed here and denoting the scaling of the fidelity as $\phi(r)$ (ie. $F(r)\gtrapprox 1-\phi(r)$) we will have that asymptotically $P(0;r)\approx \phi(r)^{\sqrt{r}}$. As mentioned above, it could be possible to obtain better exponents using higher-order decompositions. We leave this exploration to future studies. The results discussed in this section show that the QITP algorithm might not be suitable in general for long time propagation but could prove useful as a preconditioner to prepare state close to the ground-state that can then help the convergence of other algorithms that might be asymptotically more efficient (as e.g~\cite{Near}).\\

\section{Conclusions \label{sec:conclusions}}
In this work, we proposed a quantum version of the Imaginary Time Propagation algorithm for an arbitrary Hamiltonian. The method was tested and validated for two different, simple physical problems:  finding the ground state of the Hydrogen atom expanded over a STO-2G basis set, and the spin ground state of two neutrons interacting with a realistic nuclear Hamiltonian at fixed distance. We showed that a single application of the propagator over a sufficiently large imaginary time interval is sufficient to obtain a very good approximation of the ground state. 
Moreover, results of the imaginary time evolution for a generic time step and further improvements have been discussed. The robustness of the algorithm against quantum noise was tested both in a device level simulation of a transmon qudit using a single customized gate and on a real quantum computer using a standard gate set, proving the potential for solving more general problems in fields such as nuclear structure, solid state physics or quantum chemistry. 
As in the classical ITP algorithms, one of the main problems is the choice of the initial state. In particular, in the limit of $\tau\!\gg\!1$ starting from a state loosely overlapping with the ground state implies the need for a very large number of repetitions of the measurements in order to reach a reasonable accuracy since the success probability is directly proportional to the overlap probability with the GS. This can be improved by coupling the method with effective methods for preparing the initial state. Even though asymptotically more efficient methods are known (such as~\cite{Ge2019,Near}), the low ancilla requirement of the present method and the possibility to leverage Optimal Control techniques could allow QITP to be used successfully for short propagation times in order to prepare better approximations to the ground state as starting point of more efficient techniques.
Finally, this algorithm can be generalized to a complex time propagation. A study along this direction is underway.

\section{Acknowledgments}
\label{sec:Acknowledgments}
FT is supported by the Q@TN grant ANuPC-QS. PL is supported by the Q@TN grant ML-QForge. This work was prepared in part by LLNL under Contract DE-AC52-07NA27344 with support from the Laboratory Directed Research and Development grant 19-DR-005.\\

\appendix

\section{Unitarity of the ITP propagator}
\label{supp_unitary}

The operator $\hat{U}(\tau)$ in Eq.~\eqref{eq:Utau} of the main text can also be written as:

\begin{equation}
    \hat{U}(\tau)=\hat{\sigma_x} \otimes  \frac{\mathbb{1}}{\sqrt{\mathbb{1}+e^{-2\left(H-E_T\right)\tau} }  }+\hat{\sigma_z} \otimes \hat Q_{\rm ITP}(\tau).
\end{equation}
Here we show in Eq.~\eqref{eq:Utau} that the operator $U(\tau)$ is unitary.
\begin{widetext}
\begin{multline}
\hat{U}^\dagger\,\hat{U}= \left(
\begin{array}{cc}
e^{-\left(\hat{H}-E_T\right)\tau} & \mathbb{1}\\
\mathbb{1} & -e^{-\left(\hat{H}-E_T\right)\tau}\\
\end{array}\right) \frac{1}{\sqrt{\mathbb{1}+e^{-2\left(\hat{H}-E_T\right)\tau} }}  
\, \frac{\mathbb{1}}{\sqrt{\mathbb{1}+e^{-2\left(\hat{H}-E_T\right)\tau} }  } \left(
\begin{array}{cc}
e^{-\left(\hat{H}-E_T\right)\tau} & \mathbb{1}\\
\mathbb{1} & -e^{-\left(\hat{H}-E_T\right)\tau}\\
\end{array}\right)\\ 
= \frac{\mathbb{1}}{ {\mathbb{1}+e^{-2\left(\hat{H}-E_T\right)\tau} } } \left(
\begin{array}{cc}
e^{-\left(\hat{H}-E_T\right)\tau} & \mathbb{1}\\
\mathbb{1} & -e^{-\left(\hat{H}-E_T\right)\tau}\\
\end{array}\right)  \left(
\begin{array}{cc}
e^{-\left(\hat{H}-E_T\right)\tau} & \mathbb{1}\\
\mathbb{1} & -e^{-\left(\hat{H}-E_T\right)\tau}\\
\end{array}\right) \\
= \frac{\mathbb{1}}{\mathbb{1}+e^{-2\left(\hat{H}-E_T\right)\tau} } \left(
\begin{array}{cc}
\mathbb{1}+ e^{-2\left(\hat{H}-E_T\right)\tau} & 0 \\
0&\mathbb{1}+e^{-2 \left(\hat{H}-E_T\right)\tau}  \\
\end{array}\right) =  \left(
\begin{array}{cc}
\mathbb{1} & 0\\ 0&\mathbb{1}\\
\end{array}\right)\\ \label{Uni}
\end{multline}
\end{widetext}
In the proof we used the fact that the operators $\mathbb{1}$ and $e^{-\left(\hat{H}-E_T\right)\tau}$ are  Hermitian. Also, in the second line of eq.~\eqref{Uni}, we have used that the factor $\frac{1}{1+e^{-2\left(\hat{H}-E_T\right)\tau} }$ commutes with the identity and $e^{-\left(\hat{H}-E_T\right)\tau} $.
\section{Optimal control approach}
\label{supp:opt}
To implement the device level simulations presented in the main text, we mapped the physical system under investigation onto the levels  of a superconducting a three-dimensional transmon architecture~\cite{Optimalcontrol}, and solved for the control Hamiltonian $\hat H_c(t)$ in the general optimization problem:
\begin{align}
   \label{eq:constraint}
  U(\tau) = \mathcal{T}\exp{-\frac{i}{\hbar}\int_0^T \left[\hat H^{(4)}_{d}+\hat H_c(t)\right]dt}\,,
\end{align}
where $\hat H^{(4)}_d$ is the Hamiltonian for a 3D transmon coupled to a readout cavity (up to forth order in the phase across the Josephson junction), the notation $\mathcal{T} \exp$  stands for a time-ordered exponential, and T is the duration of the control pulse (for additional details, we refer the interested reader to Ref.~\cite{Optimalcontrol} ). 
Eq.~\eqref{eq:constraint} was solved by means of the Gradient Ascent Pulse Engineering (GRAPE) algorithm~\cite{GRAPE}.

The output of the quantum device is obtained in the presence of realistic quantum hardware noise using the open source quantum optics toolbox (QUTIP)~\cite{Qutip}.\\
The unitary operator $U(\tau)$ of Eq.~\eqref{eq:Utau} is a $2\,N$-by-$2\,N$ matrix (where $N$ is the total number of physical states), which we compute using the QUTIP {\em expm} function.  
The control pulse is obtained with a $99.9999\%$ fidelity using the same pulse duration ($T=50$ ns) and maximum drive strength ($20$ MHz) adopted in Ref.~\cite{Optimalcontrol}. The relaxation time was chosen as $T_1=50$ ns and the dephasing time as $T_2=7$ ns. The GRAPE code was fed with a pulse initialized using  the "ZERO" class of \textit{qutip.control.pulseoptim} (the initial pulse is a complex array of zeros), and the error fidelity threshold was set to less than $10^{-6}$.\\

\section{Nuclear Hamiltonian}
\label{supp:nuclear}
\begin{figure}[t]
	\includegraphics[scale=0.7]{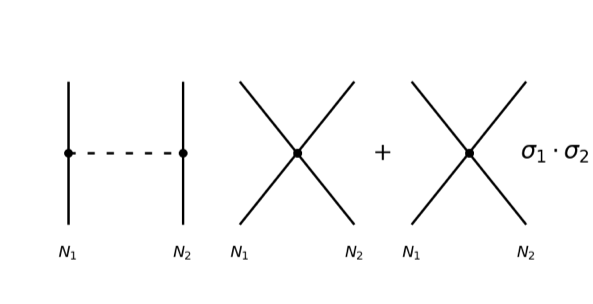}
	\caption{Schematic description of the leading order nucleon- nucleon interaction. The left diagram depicts a single pion exchange while the middle and right diagrams depict a spin- independent and spin-dependent contact term, respectively \label{fig:Fey}. } 
\end{figure}

The simplest realistic model of the nucleon-nucleon interaction can be based on the leading order (LO) of a chiral effective field theory ($\chi$-EFT). In this approximation, the nucleon-nucleon interaction is given by the sum of three terms (see Fig.~\ref{fig:Fey}). The first describes a mid range interaction due to the exchange of a pion. The second and the third are effective contact terms to be introduced in the renormalization procedure. These terms give rise to  spin-independent (SI) and spin-dependent (SD) components, so that the Hamiltonian can be expressed as $H=T+V^{\rm SI}+V^{\rm SD}$, where $T$ is the kinetic energy of the nucleons.\\
In this work we are not interested in reproducing the whole dynamics of the particles, but rather in studying the spin dynamics of two neutrons kept at fixed separation. More explicitly, considering a system of two neutrons,
we neglect the spin independent terms in $H$ and restrict ourselves to the SD interaction evaluated at a given separation $\vec{r}=\vec{r_{2}}-\vec{r_{1}}$. $V^{\rm SD}$ can then be divided into a vector and a tensor component as\\
\begin{equation}
V^{SD}=A_{1}(\vec{r}) \sum_\alpha \sigma^{1}_{\alpha} \sigma^{2}_{\alpha} + \sum_{\alpha \beta} \sigma^{1}_{\alpha} A^{\alpha\beta}_2(\vec{r})\sigma^2_{\beta}\,,
\end{equation}
where $\sigma$ are the Pauli matrices acting on the nucleon spin states. $A^1(\vec{r})$ and $A^2(\vec{r})$ can be obtained evaluating the coordinate-dependent parts of the nucleon-nucleon interaction considered (for the explicit functional forms see e.g. Refs.~\cite{37, 38}).

\section{Bounds on the success probability}
\label{app:sprob_proof}

Consider an initial state given as
\begin{equation}
\label{eq:psi_start_dec}
\ket{\Psi} = |c_0| \ket{\phi_0} + \sqrt{1-|c_0|^2} \ket{\phi_0^\perp}
\end{equation}
with $0<|c_0|\leq1$ the overlap with the ground state. The $\ket{\phi_0^\perp}$ state is the superposition of the other eigenstates of $H$. After the application of $U_0(\tau)$ to this state, the component with the ancilla in $\ket{0}$ reads
\begin{equation}
\begin{split}
\label{eq:psiinzero}
\ket{\Psi_{0}}=&\frac{e^{-\tau(E_0-E_T)}}{\sqrt{1+e^{-2\tau(E_0-E_T)}}} |c_0| \ket{\phi_0}\otimes\ket{0} \\
&+ \sqrt{1-|c_0|^2} \sum_{n>0}\frac{e^{-\tau(E_n-E_T)}}{\sqrt{1+e^{-2\tau(E_n-E_T)}}} \langle \phi_n|\phi_0^\perp\rangle \ket{\phi_n}\otimes\ket{0}\;,
\end{split}
\end{equation}
where we used the energy eigenbasis. It's then easy to show that, if we call $\Delta=E_1-E_0$ the energy gap , the success probability is bounded by
\begin{equation}
\begin{split}
P(0)\geq&\frac{e^{-2\tau(E_0-E_T)}}{1+e^{-2\tau(E_0-E_T)}}\,|c_0|^2\\
&+ (1-|c_0|^2)\frac{e^{-2\tau(\|H\|_\infty-E_T)}}{1+e^{-2\tau(\|H\|_\infty-E_T)}}\\
P(0)\leq& \frac{e^{-2\tau(E_0-E_T)}}{1+e^{-2\tau(E_0-E_T)}}\,|c_0|^2 \\
&+ (1-|c_0|^2)\frac{e^{-2\tau(E_0+\Delta-E_T)}}{1+e^{-2\tau(E_0+\Delta-E_T)}}\;,
\end{split} \label{eq:ancprob_bound}
\end{equation}
where $\|\cdot\|_\infty$ denotes the spectral norm and we have used the fact that 
\begin{equation}
f(x) = \frac{e^{-x}}{\sqrt{1+e^{-x}}}\;,
\end{equation}
is monotonically decreasing in $x$. This immediately shows that if we take $E_0-E_T\gg 1/\tau$ the success probability drops quickly to zero. One way to ensure this is not the case is to use an upperbound of $E_0$ for $E_T$, as can be found e.g. using a variational method. In order to see how bad this upperbound can be, it might be useful to rewrite the bounds as  
\begin{equation}
\begin{split}
P(0)\geq&\frac{1}{1+e^{-2\tau|E_0-E_T|}}\,|c_0|^2\\
&+ (1-|c_0|^2)\frac{e^{-2\tau(\|H\|_\infty-E_T)}}{1+e^{-2\tau(\|H\|_\infty-E_T)}}\\
P(0)\leq& \frac{1}{1+e^{-2\tau|E_0-E_T|}}\,|c_0|^2\\
&+ (1-|c_0|^2)\frac{e^{-2\tau (E_0+\Delta-E_T)}}{1+e^{-2\tau (E_0+\Delta-E_T)}}\;.
\end{split}
\end{equation}
For the worse possible upperbound $E_T=\|H\|_\infty$ and in the $\tau\to\infty$ limit we find
\begin{equation}
\label{eq:best_ps}
1\geq P(0) \geq \frac{1+|c_0|^2}{2}\;.
\end{equation}
In practice we want to choose $E_T$ closer to $E_0$ and for $E_T<\|H\|_\infty$ we find instead
\begin{equation}
1\geq P(0) \geq \,|c_0|^2\;.
\end{equation}
In the main text we performed the approximation $\|H\|_\infty-E_T\gg 1/\tau$ and used the worse lower bound
\begin{equation}
P(0)\geq\frac{1}{1+e^{-2\tau|E_0-E_T|}}\,|c_0|^2
\end{equation}
which still obtains the asymptotic limit found above.\\
In the following, we showed a real simulation of dependence of the fidelity and ancilla probability on the parameter $E_T$.\\

\section{Bounds on ground-state fidelity}
\label{app:gfid_proof}

We provide here a complete proof for the conditions to guarantee convergence to the ground-state in the large imaginary time limit of the quantum ITP presented in Sec.~\ref{sec:quantumdmc}. Let's start with the full expression of the state after the application of the $\hat{U}(\tau)$ unitary in Eq.~\eqref{eq:Utau} of the main text
\begin{equation}
\begin{split}
&\ket{\Psi(\tau)} = \frac{e^{-\tau(E_0-E_T)}}{\sqrt{1+e^{-2\tau(E_0-E_T)}}} c_0 \ket{\phi_0}\otimes\ket{0} \\
&+ \sqrt{1-|c_0|^2} \sum_{n>0}\frac{e^{-\tau(E_n-E_T)}}{\sqrt{1+e^{-2\tau(E_n-E_T)}}} \langle \phi_n|\phi_0^\perp\rangle \ket{\phi_n}\otimes\ket{0}\\
&+ \frac{c_0 \ket{\phi_0}\otimes\ket{1}}{\sqrt{1+e^{-2\tau(E_0-E_T)}}}\\
&+\sqrt{1-|c_0|^2} \sum_{n>0}\frac{\langle \phi_n|\phi_0^\perp\rangle}{\sqrt{1+e^{-2\tau(E_n-E_T)}}}  \ket{\phi_n}\otimes\ket{1}\;,
\end{split}
\end{equation}
with the (normalized) state $\ket{\phi_0^\perp}$ defined in Eq.~\eqref{eq:psi_start_dec}.
Upon a successful measurement of the reservoir qubit in $\ket{0}$, occurring with probability $P(0)$, the resulting normalized state reads
\begin{equation}
\begin{split}
&\ket{\Psi_0(\tau)} = \frac{e^{-\tau(E_0-E_T)}}{\sqrt{1+e^{-2\tau(E_0-E_T)}}} \frac{c_0}{\sqrt{P(0)}} \ket{\phi_0} \\
&+ \frac{\sqrt{1-|c_0|^2}}{\sqrt{P(0)}} \sum_{n>0}\frac{e^{-\tau(E_n-E_T)}}{\sqrt{1+e^{-2\tau(E_n-E_T)}}} \langle \phi_n|\phi_0^\perp\rangle \ket{\phi_n}\;.
\end{split}
\end{equation}
The state fidelity with the ground-state $\ket{\phi_0}$ is then
\begin{equation}
F(\tau)=\left|\langle\phi_0|\Psi_0(\tau)\right|^2 = \frac{e^{-2\tau(E_0-E_T)}}{1+e^{-2\tau(E_0-E_T)}} \frac{|c_0|^2}{P(0)}\;. \label{eq:fidelityGS}
\end{equation}
This expression can be bounded from above and below using the bounds on the success probability found earlier in Appendix~\ref{app:sprob_proof}. In particular we have the lower bound
\begin{equation}
F(\tau)\geq \left(1+\frac{1-|c_0|^2}{|c_0|^2}\frac{1+e^{2\tau(E_0-E_T)}}{1+e^{2\tau(E_0+\Delta-E_T)}}\right)^{-1}\;.\label{eq:bound_fidelityGS}
\end{equation}
Using this last equation one can prove that fidelity is always greater than the initial one.
We can now distinguish two situations
\begin{itemize}

 \item for $E_T\geq E_0+\Delta$ the lower bound becomes
\begin{equation}
\label{eq:fidelity_large_et}
 F(\tau)\geq \left(1+\frac{1-|c_0|^2}{|c_0|^2}\right)^{-1}=|c_0|^2\;.
\end{equation}
This is manifestly greater than the initial fidelity $F(0)=|c_0|^2$ and the quantum ITP algorithm is guaranteed not to worsen the ground-state fidelity for non-zero values of $\tau$
\item for $E_0+\Delta>E_T>E_0$ the lower bound is instead
\begin{equation}
 F(\tau)\geq \left(1+2\frac{1-|c_0|^2}{|c_0|^2}e^{-2\tau|E_0+\Delta-E_T|}\right)^{-1}\;,
\end{equation}
which, for long propagation times $\tau\gg1/(E_0+\Delta-E_T)$, converges to $1$ exponentially fast.
\end{itemize}
As the derivation above shows, for a good convergence of the quantum ITP algorithm described in the main text it is important to choose $E_0+\Delta>E_T>E_0$. This shows that, if we use a variational simulation to get an upperbound on $E_0$, we also need to ensure it's accuracy is better than the energy gap $\Delta$. For many problems where the ground-state energy is not known with precision better than $\Delta$, we cannot ensure the algorithm will correctly converge to the ground-state. Nevertheless, the quantum ITP scheme can still be useful in situations where both $E_0$ and $\Delta$ are known beforehand and the task is only to prepare a good approximation to the ground-state.

\begin{figure}[t]
\centering
    \includegraphics[scale=0.2]{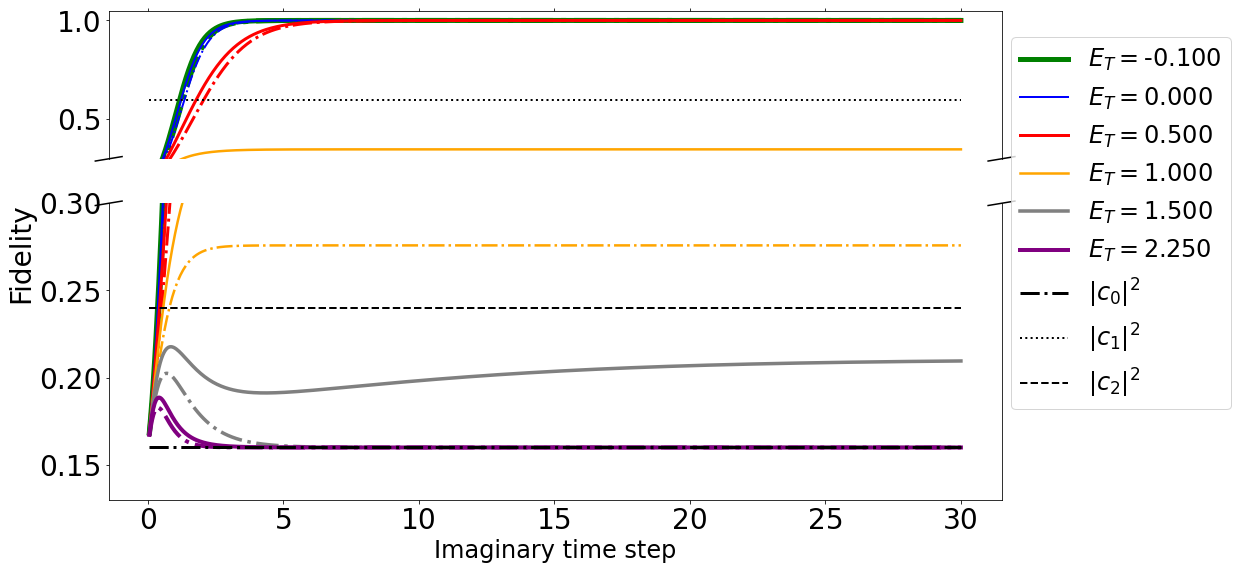}
    \caption{Fidelity with GS for different values of $E_T$. Lines represent the real fidelities, the dashed lines the lower bounds. The initial fidelity between the eigenstates are shown by the horizontal lines ($c_0$,$c_1$ and $c_2$) } 
     \label{fig:fidelity_ET}
\end{figure}

\begin{figure}[t]
    \centering
    \includegraphics[scale=0.25]{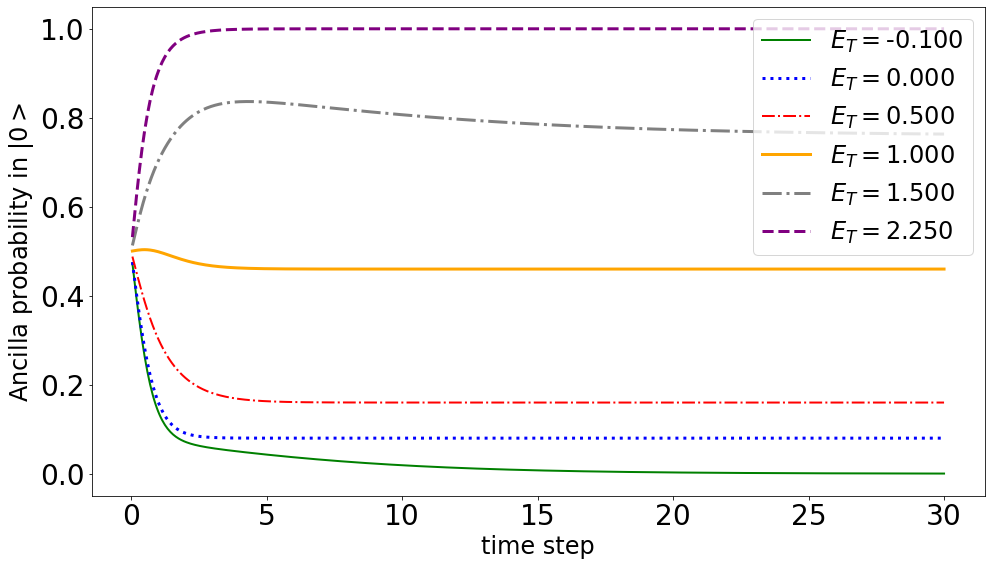}
    \caption{Ancilla probability for different values of $E_T$}
    \label{fig:prob_ancilla_ET}
\end{figure}

Using the previous equations we tested how the fidelity  and probability of measuring the ancilla in $\ket{0}$ changed for different values of $E_T$ considering as system a simple one whose spectra is given by $[0,\,1,\,\frac{\pi}{2}]$.  We computed the exact fidelity with GS using eq.~\eqref{eq:fidelityGS} and lower bound limit given by eq.~\eqref{eq:bound_fidelityGS} and our results are shown in Fig.~\ref{fig:fidelity_ET}.  Fig.~\ref{fig:prob_ancilla_ET} show the ancilla probability.\\
For all the values of $E_T$, one can observe that the normalized final state is closer to to GS than the initial one because the final fidelity with GS is greater than initial one. Nevertheless, for trial energy $E_T\!>\!E_1$, in particular in the limit  $\tau\!\gg\! 1$, the fidelity with GS asymptotically goes to a value equal or greater than initial one. For $E_T\!<\!E_1$, applying the quantum ITP algorithm one reach the ground state.\\
Looking instead the ancilla probability plot one see the contrary behaviour. For $E_T<E_0$, it drops exponentially to the 0 value. In this case, the final unitary gate $U(\tau)$ eventually reduces to an $X$ gate on the ancilla qubit, implying that the final probability of measuring the ancilla in the $\ket{0}$ state tends to 0. Moreover, in a realistic simulations, the measured probability of the $\ket{0}$ state of the ancilla would exclusively be the result of quantum noise of machine. In the other case, $E_T \ge E_0$ one can demostrate the ancilla probability has lower bound given by eq.~\eqref{eq:ancprob_bound}.\\
In conclusion, one may choose as optimal parameter for $E_T$, a value that is between the ground state $E_0$ and the energy of first excited state $E_1$ because maximizes the overlap with GS and probability of measuring the ancilla in $\ket{0}$.

\section{Partial tomography: qubit estimation}\label{app:tomography}
In this appendix we provide details on the tomography procedure employed in Sec.~\ref{sec:Evolution} of the main text. This is different from standard qubit tomography in that we perform measurements of $\langle X\rangle$, $\langle Y\rangle$ and $\langle Z\rangle$ but directly fit the results to a pure state.\\
Our aim is to evaluate  $P_0$, $P_1$ and $\theta_r=\theta_1-\theta_0$ of the following qubit:
\begin{equation}
    \psi=\left(\begin{array}{c}
    a+i\,b\\
    c+i\,d
    \end{array}\right)=e^{i\theta_0}\;\left(\begin{array}{c}
    \sqrt{P_0} \\
    \sqrt{P_1} e^{-i (\theta_1-\theta_0)}
    \end{array}\right)\,,
\end{equation}
where we can  neglect $\theta_0$ because it represents a global phase of  qubit.
We used the following algorithm:
\begin{enumerate}
\item Measure the probabilities of the bare circuit (we have $P_0$ and $P_1$)
\item Measure the probabilities of the bare circuit with a final $R_y(-\frac{\pi}{2})$ rotation.

\item Measure the probabilities of the bare circuit with a final $R_x(-\frac{\pi}{2})$ rotation

\end{enumerate}
We need to the two rotations to establish uniquely the relative phase. In particular, one finds the probability of the $\ket{0}$ with a $R_y(-\frac{\pi}{2})$ gate is given by
\begin{equation}
\begin{split}
P_0^y &=\frac{1}{2} \left((a+c)^2+(b+d)^2\right)\\
&=\frac{1}{2}\left( (a^2+b^2+c^2+d^2)+2 ac+2 bd\right)\\
&= \frac{1}{2} \left( 1+2 \sqrt{P_0 P_1} \cos(\theta_0) \cos(\theta_1)+2 \sqrt{P_0 P_1} \sin(\theta_0) \sin(\theta_1) \right)\,.
\end{split}    
\end{equation}
Therefore, we have:
\begin{equation}
\cos(\theta_0-\theta_1)= \frac{1}{\sqrt{P_0 P_1} } \left(P^y_0-\frac{1}{2}(P_0+P_1)\right)\,.  \label{eq:tomo_cos}
\end{equation}
Similarly for the $R_x$ rotation we obtain:
\begin{equation}
\sin(\theta_0-\theta_1)= \frac{1}{\sqrt{P_0 P_1} } \left(P^x_0-\frac{1}{2}(P_0+P_1)\right)\,. \label{eq:tomo_sin}
\end{equation}
With these two equations we can establish unequivocally the angle $\theta_0-\theta_1$. 

\subsection{Density matrix formulation}
The general density matrix for a single qubit can be expressed as follows:
\begin{equation}
    \rho=\frac{1}{2}(1+c_x\,X+c_y\,Y+c_z\,Z)\,.
\end{equation}
Applying a $R_x(-\frac{\pi}{2})$ and $R_x(-\frac{\pi}{2})$ to $\rho$, one gets:
\begin{equation}
    \rho_x= R_x(-\frac{\pi}{2}) \rho R_x(-\frac{\pi}{2})^\dagger= \frac{1}{2}(1+c_x\,X -c_y\,Z+c_z\, Y)
\end{equation}
\begin{equation}
    \rho_y= R_y(-\frac{\pi}{2}) \rho R_y(-\frac{\pi}{2})^\dagger= \frac{1}{2}(1+c_x\,Z +c_y\,Y-c_z\, X)\,.
\end{equation}
Measuring the $\ket{0}$ state for both the two density matrices, one obtains:
\begin{equation}
\begin{split}
    P^x_0=&\frac{1}{2}(1-c_y)\\
    P^y_0=&\frac{1}{2}(1+c_x)\,.
\end{split}
\end{equation}

Using the formula in eqs.~\eqref{eq:tomo_cos} and \eqref{eq:tomo_sin} in the case of a general density matrix, one gets
\begin{equation}
    \begin{split}
    \theta_1-\theta_0&=-\arctan\left(\frac{\left(P^x_0-\frac{1}{2}(P_0+P_1)\right)}{\left(P^y_0-\frac{1}{2}(P_0+P_1)\right)} \right) \\
    & =- \arctan\left( \frac{\frac{1}{2}(1-c_y)-\frac{1}{2}}{\frac{1}{2}(1+c_x)-\frac{1}{2}}\right)\\
    & =\arctan\left( \frac{c_y}{c_x}\right)\,.
\end{split}
\end{equation}

In the following, we show how its work with a qubit state, $\ket{\psi}=\alpha \ket{0}+\beta \ket{1}$, using the Bloch-Redfield density matrix~\cite{Bloch_Redfield} shown in eq.~\eqref{eq:Bloch_ReF}.
\begin{equation}
    \rho=\left( \begin{array}{cc} 1+(\left|\alpha\right|^2-1) e^{-\frac{t}{T_1}} & \alpha\beta^* e^{-\frac{t}{T_2}}\\ \alpha^*\beta e^{-\frac{t}{T_2}} & |\beta|^2 e^{-\frac{t}{T_1}}\\
    \end{array}\right)\label{eq:Bloch_ReF}
\end{equation}
Table~\ref{tab:tomography} shows our expactation values of $\langle X \rangle$ , $\langle Y \rangle$ and$\langle Z \rangle$ using the exposed tomography method with the Bloch Redfield density matrix (eq.~\eqref{eq:Bloch_ReF}). for the following randomic state,
\begin{equation}
\ket{\psi}= (0.1792) \ket{0} +  (0.8741-0.4515\, i) \ket{1} \,.
\end{equation}
We observe that our purification and tomography procedure resist to the dephasing error, but not to the relaxation process.

\begin{table}[t]
    \centering
    \begin{tabular}{|c|c|c|c|}
         \hline
         & $\langle X \rangle$ & $\langle Y \rangle$& $\langle Z \rangle$ \\ [0.1cm]\hline 
         without noise  & 0.3132 &-0.1618 &-0.9358\\[0.25cm]

         $\frac{t}{T_2}=1$ and $T_1=0$   &  0.3132& -0.1618&-0.9358\\[0.25cm]
 $\frac{t}{T_1}=\frac{t}{T_1}=1$   
         & 0.8508 & -0.4395&0.2879\\[0.1cm]
         \hline
    \end{tabular}
    \caption{Results of $\langle X \rangle$ , $\langle Y \rangle$ and$\langle Z \rangle$ using the exposed tomography method with the Bloch Redfield density matrix using a randomic qubit state}
    \label{tab:tomography}
\end{table}

\section{Discussion of the results obtained on the IBM system} 
\label{supp:IBM_Discussion}

We tested the validity of the method presented in this paper running the code in Fig.~\ref{gate_Hyl} of the main text on different free access IBM Quantum Experience machines~\cite{IBM_quantum_exp}.\\
All tests were performed using 8192 shots starting from the $\ket{x}=\frac{1}{\sqrt{2}} \left( \ket{0}+\ket{1} \right) $ state. The initial fidelity with the GS is $0.361$, the trial energy $E_T=E_0$ and $\tau\,=\,15$ Hartee$^{-1}$. The numerical results are shown in Table~\ref{tab:ibm} for different QPUs. The errors of the QPUs were reported in the same table.

\begin{table*}[t]
    \centering
    \begin{tabular}{|c|c|cc|ccc|cccc|}
        \hline
        QPUs & GS Fidelity &  R. $q_0$&  R. $q_1$ & Cnot 
        &  $\sqrt{\sigma_x}(q_0)$  &$\sqrt{\sigma_x}(q_1)$ &$T_1(q_0)\,(\mu s)$ &$T_2(q_0)\,(\mu s)$ &$T_1(q_1)\,(\mu s)$ &$T_2(q_1)\,(\mu s)$\\
        \hline 
$ibmq$\textunderscore$lima$ &0.885(9)& 0.0207& 0.0169&  0.0051& 0.00021&  0.00023& 71.88&  116.87& 98.84&  126.62\\
$ibmq$\textunderscore$manila$ &
0.882(14)&  0.020&  0.023& 0.0062&  0.00016& 0.00017&  184.94&  131.68&  100.90&  97.51 \\
$ibmq$\textunderscore$santiago$ &
0.950(3)& 0.014 & 0.015&0.0072& 0.00023&0.00022&113.05& 64.23& 202.74& 80.14 \\
$ibmq$\textunderscore$quito$ &
0.889(4)& 0.036& 0.018& 0.017& 0.00077& 0.00031& 78.81& 67.26& 114.10& 154.79\\
\hline
\end{tabular}
\onecolumngrid
\caption{Obtained fidelities between GS and the normalized state obtained after quantum ITP simulation with $\tau=15.0$ Hartee$^{-1}$ for different IBM QPUs. The parameters (Readout error of qubit 0 and 1, of Cnot and $\sqrt{\sigma_x}$ implementations) of IBM QPUs are also shown. The initial fidelity was  $0.361$}
    \label{tab:ibm}
\end{table*}

\section{Proofs of Trotter decomposition \label{app:Trotter_proof}}
A simple generalization of the QITP unitary from eq.~\eqref{eq:Utau} is given by the following operator:
\begin{equation}
    \hat{U}(\tau,\eta)\,=\,\begin{pmatrix}
    \frac{e^{-\tau (\hat{H}-E_T)}}{\sqrt{\eta^2+e^{-2\tau(\hat{H}-E_T)} }} &   \frac{\eta}{\sqrt{\eta^2+e^{-2\tau(\hat{H}-E_T)} }}\\
    \frac{\eta}{\sqrt{\eta^2+e^{-2\tau(\hat{H}-E_T)} }} & -\frac{e^{-\tau (\hat{H}-E_T)}}{\sqrt{\eta^2+e^{-2\tau(\hat{H}-E_T)} }}
    \end{pmatrix}\,,
\end{equation}
where $\eta$ is a real positive number. We recover the standard operator $\hat{U}(\tau)$ of eq.~\eqref{eq:Utau} when $\eta=1$.\\
We are interested in implementing the Trotter decomposition for the evolution. We apply $r$ times a short-time $\hat{U}(\tau,\eta)$ where at implementation we measure the ancilla and we obtain the ancilla in $\ket{0}$ with some probability. The whole operator would be obtained from:
\begin{equation}
    \left(\hat{M_0} \hat{U}(\tau,\eta) \right)^r = \begin{pmatrix}
    \left( \frac{e^{-\tau (\hat{H}-E_T)}}{\sqrt{\eta^2+e^{-2\tau(\hat{H}-E_T)} }}\right)^r & 0 \\
    0 & 0\\
    \end{pmatrix}\,,
\end{equation}
where $\hat{M_0}=\rvert0\rangle\langle0\lvert$ is the projector operator to the $\ket{0}$ state of the ancilla qubit.

We arrive to the following Lemma. The case for  $\eta=1$ is the Lemma of the main text.\\
\begin{remark}
Let $H$ be a Hermitian operator expressed as the sum of $L$ Hermitian operators $\{\hat{H_1},..,\hat{H_L}\}$ as $\hat{H}=\sum_l^L \hat{H_l}$ and $\Lambda_T=\,\max_j\,\left\|\hat{H}_j-E_T\right\|_{\infty}$. Then, 
\begin{equation}
\label{eq:app_lemma}
\left\|\hat{Q}_{ITP}(\eta,\tau)-\prod_k^L\hat{Q}^{(k)}_{ITP}(\eta,\tau)\right\|\leq L^2\Lambda_T^2 \tau^2\,,
\end{equation}
where
\begin{equation}
\hat{Q}_{ITP}^{(k)}(\eta,\tau)\,=\,  \frac{e^{-\tau \left(\hat{H_k}-\frac{E_T}{L}\right)}}{\sqrt{\eta^2+e^{-2\tau\left(\hat{H_k}-\frac{E_T}{L} \right)} }};.
\end{equation}
\end{remark}

\begin{proof}
We begin the discussion by generalizing the result in appendix F of Ref.~\cite{Trotter_error} to the QITP operator. We first introduce the notation $\mathcal{R}_k(f)$ to denote the remainder of the truncated Taylor series of an analytic function $f$. Consider the following function, a simplified form of the $QITP$ operator,
\begin{equation}
f(x) = \frac{e^{-x}}{\sqrt{\eta^2+e^{-x}}}\;,
\end{equation}
with $\eta>0$. This is analytic for any real $x$ and using either Langrange's expression for the Taylor series remainder we can write
\begin{equation}
\begin{split}
\mathcal{R}_k(f;x) &= \sum_{m=k+1}^\infty \frac{f^m(x)}{m!}x^m= \frac{f^{k+1}(\xi)}{(k+1)!}x^{k+1}\\
\end{split}
\end{equation}
for some $\xi\in[0,x]$. In the following we need a bound for $k=1$ and we have
\begin{equation}
\begin{split}
\label{eq:deriv_bound}
f^2(x) &= \frac{e^{-x}}{\sqrt{\eta^2+e^{-x}}}-4\frac{e^{-3x}}{\left(\eta^2+e^{-x}\right)^{3/2}}+3\frac{e^{-5x}}{\left(\eta^2+e^{-x}\right)^{5/2}}\\
&=\frac{e^{-x}}{\sqrt{\eta^2+e^{-x}}}\left(1-\frac{e^{-2x}}{\eta^2+e^{-2x}}\left(4-3\frac{e^{-2x}}{\eta^2+e^{-2x}}\right)\right)\\
&\leq\frac{e^{-x}}{\sqrt{\eta^2+e^{-x}}}\left(1-\frac{e^{-2x}}{\eta^2+e^{-2x}}\right)\\
&\leq\frac{e^{-x}}{\sqrt{\eta^2+e^{-x}}}\leq 1\;.
\end{split}
\end{equation}
Consider now the operator function
\begin{equation}
\begin{split}
f(t\hat{H}) &= \sum_n \rvert n\rangle\langle n\lvert f(tE_n)\\
&=\sum_{m=0}^\infty\sum_n \rvert n\rangle\langle n\lvert \frac{f^m(tE_n)}{m!}(tE_n)^m \,,
\end{split}
\end{equation}
where $\ket{n}$ indicates the eigenstate of $H$ with eigenvalue $E_n$.\\
Noting that $f(x)$ is positive definite we can bound the remainder for $k=1$ with
\begin{equation}
\left\|\mathcal{R}_1(f,t\hat{H})\right\| = \left\|\frac{f^2(t\hat{H})}{2}(t\hat{H})^2\right\|\leq \frac{t^2\|\hat{H}\|^2}{2}
\end{equation}
Consider now the remainder of the product
\begin{equation}
f_L(t,\vec{x})=\prod_{l=1}^L f(tx_l)
\end{equation}
written in terms of derivatives with respect to the scalar variable $t$ (for bookkeeping purposes) 
\begin{equation}
f^1_L(t,\vec{x}) = \sum_{m=1}^L x_mf^1(tx_m) \prod_{l\neq m}^L f(tx_l)
\end{equation}
and
\begin{equation}
\begin{split}
f^2_L(t,\vec{x}) &= \sum_{m=1}^L x_m^2f^2(tx_m) \prod_{l\neq m}^L f(tx_l)\\
&+\sum_{m=1}^L\sum_{k\neq l}^L x_mx_k\,f^1(tx_m)f^1(tx_k) \prod_{l\neq m,k}^L f(tx_l)\\
&\leq \sum_{m=1}^L x_m^2f^2(tx_m) +\sum_{m=1}^L\sum_{k\neq l}^L x_mx_kf^1(tx_m)f^1(tx_k)\\
&\leq L^2\max_k[x_k^2]\;,
\end{split}
\end{equation}
since $|f^1(x)|\leq1$ and $f^2(x)\leq1$ for any real $x$ following the same argument used to derive eq.~\eqref{eq:deriv_bound}. We now have for the remainder
\begin{equation}
\begin{split}
\mathcal{R}_1\left[\prod_{l=1}^L f(t\hat{H_l})\right] &= \left\|\frac{f^2_L(t,\hat{H})}{2}t^2\right\| \\
&\leq \frac{t^2L^2\max_l[\|\hat{H}\|^2_l]}{2}\;.
\end{split}
\end{equation}

In order to simplify the notation we will use $Q=Q_{ITP}(\eta,\tau)$ and $Q^{(k)}=Q^{(k)}_{ITP}(\eta,\tau)$ whenever there is no risk of confusion.
Using the triangular inequality, we get:
\begin{equation}
\begin{split}
\left\|Q-\prod_k^LQ^{(k)}\right\|&=\left\|\mathcal{R}_1\left(Q-\prod_k^LQ^{(k)}\right)\right\|\\
&=\left\|\mathcal{R}_1\left(Q\right)-\mathcal{R}_1\left(\prod_k^LQ^{(k)}\right)\right\|\\
&\leq\left\|\mathcal{R}_1\left(Q\right)\right\|+\left\|\mathcal{R}_1\left(\prod_k^LQ^{(k)}\right)\right\|\\
\label{eq:bound_trotter1}
\end{split}
\end{equation}
Using the results obtained above we have
\begin{equation}
\begin{split}
\left\|\mathcal{R}_1\left(Q_{ITP}(\eta,\tau)\right)\right\|&\leq \frac{\tau^2\|\hat{H}-E_T\|^2}{2}\\
&\leq\frac{\tau^2L^2\Lambda_T^2}{2}\;,
\end{split}
\end{equation}
while for the product
\begin{equation}
\left\|\mathcal{R}_1\left(\prod_k^LQ^{(k)}(\eta,\tau)\right)\right\|\leq\frac{\tau^2L^2\Lambda_T^2}{2}\;.    
\end{equation}
Note that neither bound depends explicitly on the choice of the $\eta$ parameter. The results easily follows by summing these two contributions.

\end{proof}

In order to use this to approximate the $r$ steps with total error $\epsilon$ we then need
\begin{equation}
\delta\tau = \mathcal{O}\left(\sqrt{\frac{\epsilon}{r}}\frac{1}{L\Lambda_T}\right)\ll1\;.
\end{equation}
Applying $r$ times the $\hat{U}(\delta\tau,\eta)$ with the measure of ancilla qubit in  $\ket{0}$ state, one gets that the fidelity with GS becomes using the same formula of eq. \eqref{eq:fidelityGS}:

\begin{equation}
F= \left(1+\sum_n \frac{|c_n|^2}{|c_0|^2} \frac{\left(\eta^2e^{2\,\delta\tau\,(E_0-E_T)}+1\right)^r}{ \left(\eta^2e^{2\,\delta\tau\,r\,(E_n-E_T)}+1 \right)^r } \right)^{-1}\\
\end{equation}

Using the result of eq.~\eqref{eq:bound_fidelityGS} for the fidelity bound, in the general case of $\eta$ and $r\in\mathbb{N}$ we get:
\begin{equation}
F\geq\left(1+\frac{1-|c_0|^2}{|c_0|^2}\left(\frac{\eta^2e^{2\delta\tau(E_0-E_T)}+1}{\eta^2e^{2\delta\tau(E_1-E_T)}+1}\right)^r\right)^{-1}.
\end{equation}
In order to bound the behavior in the parenthesis we use
\begin{equation}
\begin{split}
\frac{\eta^2e^{2\delta\tau(E_0-E_T)}+1}{\eta^2e^{2\delta\tau(E_1-E_T)}+1} &= \frac{\eta^2+e^{-2\delta\tau(E_0-E_T)}}{\eta^2e^{2\delta\tau\Delta}+e^{-2\delta\tau(E_0-E_T)}}\\
&=1-\eta^2\frac{e^{2\delta\tau\Delta}-1}{\eta^2e^{2\delta\tau\Delta}+e^{-2\delta\tau(E_0-E_T)}}\;.
\end{split}
\end{equation}
Assume now we choose $E_1>E_T>E_0$, this means $E_T-E_0<\Delta$ and we have
\begin{equation}
\begin{split}
\frac{\eta^2e^{2\delta\tau(E_0-E_T)}+1}{\eta^2e^{2\delta\tau(E_1-E_T)}+1}&< 1-\frac{\eta^2}{1+\eta^2}\frac{e^{2\delta\tau\Delta}-1}{e^{2\delta\tau\Delta}}\\
&\leq1-\frac{\eta^2}{1+\eta^2}\frac{2\delta\tau\Delta}{1+2\delta\tau\Delta}\;.
\end{split}
\end{equation}

Considering a small enough error $\epsilon$ or large $r$ to be in the limit $2\delta\tau\,\Delta\,<\,1$, one can prove that 
\begin{multline}
\begin{split}
\label{eq:fid_fac_bound}
\left(\frac{\eta^2e^{2\delta\tau(E_0-E_T)}+1}{\eta^2e^{2\delta\tau(E_1-E_T)}+1}\right)^r
&\leq \left(1-\frac{\eta^2}{1+\eta^2}\frac{2\delta\tau\Delta}{1+2\delta\tau\Delta}\right)^r\\
&\leq\left(1-\frac{\eta^2}{1+\eta^2}\delta\tau\Delta\right)^r\\ 
&\leq\exp\left(-\frac{\eta^2}{1+\eta^2}r\delta\tau\Delta\right)\\
\end{split}\\
=\,\mathcal{O}\left(\exp\left(-\frac{\eta^2}{1+\eta^2}\frac{\Delta}{L\Lambda_T}\sqrt{r\epsilon}\right)\right) \,.
\end{multline}
The result shown in the main text is obtained for $\eta=1$.\\
Finally we can compute the success probability as a function of $r$. Using the result found above and in Appendix~\ref{app:sprob_proof} we have
\begin{equation}
\begin{split}
P(0;r)&=\langle\Psi\lvert \left(\hat{M_0} \hat{U}(\delta\tau,\eta) \right)^{2r}\rvert\Psi\rangle\\
&=\langle\Psi\lvert \hat{Q}_{ITP}(\delta\tau)^{2r}\rvert\Psi\rangle\\
&\geq|c_0|^2 \langle\phi_0\lvert \hat{Q}_{ITP}(\delta\tau)^{2r} \rvert\phi_0\rangle\\
&\geq \frac{|c_0|^2}{(\eta^2e^{2\delta\tau(E_0-E_T)}+1)^r}
\end{split}
\end{equation}
which, for the choice $E_1>E_T>E_0$ can be rewritten more clearly as
\begin{equation}
\begin{split}
\label{eq:prob0r_lowb_app}
P(0;r)\geq \frac{|c_0|^2}{(\eta^2e^{-2\delta\tau|E_0-E_T|}+1)^r}\geq \frac{|c_0|^2}{(\eta^2+1)^r}
\end{split}
\end{equation}
which decays exponentially with the number of steps. By tuning $\eta$ one can make also this behavior to be sub-exponential, at the expense of a slower convergence of the fidelity with step number $r$. For $\eta=1$ we obtain the result shown in the main text. This results holds only for the exact short-time unitary $\hat{U}(\delta\tau,\eta)$ but using the approximation in Eq.~\eqref{eq:app_lemma} of the Lemma above we still have that the difference in probabilities is bounded as
\begin{equation}
\begin{split}
\delta P &= \left|\langle\Psi\lvert \hat{Q}_{ITP}(\delta\tau)^{2r}\rvert\Psi\rangle - \langle\Psi\lvert \prod_k^L\hat{Q}^{(k)}_{ITP}(\eta,\tau)\rvert\Psi\rangle\right|\\
&= \left|\langle\Psi\lvert \left(\hat{Q}_{ITP}(\delta\tau)^{2r} -  \prod_k^L\hat{Q}^{(k)}_{ITP}(\eta,\tau)\right)\rvert\Psi\rangle\right|\\
&\leq \left\|\hat{Q}_{ITP}(\delta\tau)^{2r} -  \prod_k^L\hat{Q}^{(k)}_{ITP}(\eta,\tau)\right\|\\
&\leq 2rL^2\Lambda_T^2\delta\tau^2\;.
\end{split}
\end{equation}
In other words, if we use an approximation for the full sequence of $r$ steps with error bounded by $\epsilon$, the success probability will be at most $2\epsilon$ smaller. Since the decay of $P(0;r)$ is exponential in $r$, the lower bound in Eq.~\eqref{eq:prob0r_lowb_app} will go to zero at some finite number of steps. This is possibly a consequence of the looseness of the bound for $\delta P$ given above.
\end{document}